\documentclass[sigconf]{acmart}
\renewcommand\footnotetextcopyrightpermission[1]{} % removes footnote with conference information in first column

\usepackage{svg}
\usepackage{algorithm}
\usepackage{algorithmic}
\usepackage{amsmath}
\usepackage{amssymb}
\usepackage{xcolor}

\usepackage{tcolorbox}
\usepackage{xcolor}
\usepackage{setspace}
\usepackage{listings}
\tcbuselibrary{listings}
\definecolor{headergray}{RGB}{90,90,90}
\definecolor{highlightorange}{RGB}{230,120,20}
\definecolor{highlightblue}{RGB}{30,90,180}
\definecolor{frameblue}{RGB}{45,70,120}

\definecolor{lightgray}{RGB}{245,245,245}

\AtBeginDocument{%
  }

\begin{document}

%%
%% The "title" command has an optional parameter,
%% allowing the author to define a "short title" to be used in page headers.
% \title{AMEM4Rec: Agentic Recommendation with Evolving Memory-Guided Collaborative Filtering}
\title{AMEM4Rec: Leveraging Cross-User Similarity for Memory Evolution in Agentic LLM Recommenders}

%%
%% The "author" command and its associated commands are used to define
%% the authors and their affiliations.
%% Of note is the shared affiliation of the first two authors, and the
%% "authornote" and "authornotemark" commands
%% used to denote shared contribution to the research.
\author{Minh-Duc Nguyen}
% \authornote{Both authors contributed equally to this research.}

% \orcid{1234-5678-9012}
% \author{G.K.M. Tobin}
% \authornotemark[1]
% \email{webmaster@marysville-ohio.com}
\affiliation{%
  \institution{Center for AI Research, VinUniversity}
  \city{Hanoi}
  % \state{Ohio}
  \country{Vietnam}
}
\email{duc.nm2@vinuni.edu.vn}

\author{Hai-Dang Kieu}

\affiliation{%
  \institution{College of Engineering and Computer
Science, VinUniversity}
  \city{Hanoi}
  \country{Vietnam}}
\email{dangkh@vinuni.edu.vn}

\author{Dung D. Le}
\affiliation{%
  \institution{College of Engineering and Computer
Science, VinUniversity}
  \city{Hanoi}
  \country{Vietnam}
}
\email{dung.ld@vinuni.edu.vn}

% \author{Aparna Patel}
% \affiliation{%
%  \institution{Rajiv Gandhi University}
%  \city{Doimukh}
%  \state{Arunachal Pradesh}
%  \country{India}}

% \author{Huifen Chan}
% \affiliation{%
%   \institution{Tsinghua University}
%   \city{Haidian Qu}
%   \state{Beijing Shi}
%   \country{China}}

% \author{Charles Palmer}
% \affiliation{%
%   \institution{Palmer Research Laboratories}
%   \city{San Antonio}
%   \state{Texas}
%   \country{USA}}
% \email{cpalmer@prl.com}

% \author{John Smith}
% \affiliation{%
%   \institution{The Th{\o}rv{\"a}ld Group}
%   \city{Hekla}
%   \country{Iceland}}
% \email{jsmith@affiliation.org}

% \author{Julius P. Kumquat}
% \affiliation{%
%   \institution{The Kumquat Consortium}
%   \city{New York}
%   \country{USA}}
% \email{jpkumquat@consortium.net}

%%
%% By default, the full list of authors will be used in the page
%% headers. Often, this list is too long, and will overlap
%% other information printed in the page headers. This command allows
%% the author to define a more concise list
%% of authors' names for this purpose.
\renewcommand{\shortauthors}{Minh-Duc et al.}

%%
%% The abstract is a short summary of the work to be presented in the
%% article.
\begin{abstract}
 % Recent efforts to integrate Large Language Models (LLMs) into recommender systems (RS) face several challenges. Fine-tuning LLMs for recommendation tasks is computationally expensive and inefficient in terms of parameter optimization. Existing studies also struggle to control LLM outputs to ensure task-aligned recommendations. Although prompt optimization avoids retraining LLMs, it suffers from context-length limitations and increases the risk of hallucination. Furthermore, LLM-based recommenders primarily exploit semantic knowledge while neglecting collaborative signals that are essential in traditional recommendation. To address these issues, we propose AMEM4Rec, an innovative agentic recommender system that leverages the internal knowledge of LLMs and integrates evolving memory-guided collaborative filtering to enhance reasoning and recommendation accuracy.
 % Recent efforts to integrate Large Language Models (LLMs) into recommender systems face persistent challenges. Fine-tuning LLMs is computationally expensive and parameter-inefficient. Prompt-based methods suffer from context-length limits and increased hallucination risk, while most approaches exploit semantic knowledge but neglect collaborative filtering (CF) signals essential for implicit preference modeling. 
 Agentic systems powered by Large Language Models (LLMs) have shown strong potential in recommender systems but remain hindered by several challenges. Fine-tuning LLMs is parameter-inefficient, and prompt-based agentic reasoning is limited by context length and hallucination risk. Moreover, existing agentic recommendation systems predominantly leverages semantic knowledge while neglecting the collaborative filtering (CF) signals essential for implicit preference modeling. To address these limitations, we propose AMEM4Rec, an agentic LLM-based recommender that learns collaborative signals in an end-to-end manner through cross-user memory evolution. AMEM4Rec stores abstract user behavior patterns from user histories in a global memory pool. Within this pool, memories are linked to similar existing ones and iteratively evolved to reinforce shared cross- user patterns, enabling the system to become aware of CF signals without relying on a pre-trained CF model. Extensive experiments on Amazon and MIND datasets show that AMEM4Rec consistently outperforms state-of-the-art LLM-based recommenders, demonstrating the effectiveness of evolving memory-guided collaborative filtering.
\end{abstract}

%%
%% The code below is generated by the tool at http://dl.acm.org/ccs.cfm.
%% Please copy and paste the code instead of the example below.
%%
% \begin{CCSXML}
% <ccs2012>
%  <concept>
%   <concept_id>00000000.0000000.0000000</concept_id>
%   <concept_desc>Do Not Use This Code, Generate the Correct Terms for Your Paper</concept_desc>
%   <concept_significance>500</concept_significance>
%  </concept>
%  <concept>
%   <concept_id>00000000.00000000.00000000</concept_id>
%   <concept_desc>Do Not Use This Code, Generate the Correct Terms for Your Paper</concept_desc>
%   <concept_significance>300</concept_significance>
%  </concept>
%  <concept>
%   <concept_id>00000000.00000000.00000000</concept_id>
%   <concept_desc>Do Not Use This Code, Generate the Correct Terms for Your Paper</concept_desc>
%   <concept_significance>100</concept_significance>
%  </concept>
%  <concept>
%   <concept_id>00000000.00000000.00000000</concept_id>
%   <concept_desc>Do Not Use This Code, Generate the Correct Terms for Your Paper</concept_desc>
%   <concept_significance>100</concept_significance>
%  </concept>
% </ccs2012>
% \end{CCSXML}

% \ccsdesc[500]{Do Not Use This Code~Generate the Correct Terms for Your Paper}
% \ccsdesc[300]{Do Not Use This Code~Generate the Correct Terms for Your Paper}
% \ccsdesc{Do Not Use This Code~Generate the Correct Terms for Your Paper}
% \ccsdesc[100]{Do Not Use This Code~Generate the Correct Terms for Your Paper}

%%
%% Keywords. The author(s) should pick words that accurately describe
%% the work being presented. Separate the keywords with commas.
%% A "teaser" image appears between the author and affiliation
%% information and the body of the document, and typically spans the
% %% page.
% \begin{teaserfigure}
%   \includegraphics[width=\textwidth]{sampleteaser}
%   \caption{Seattle Mariners at Spring Training, 2010.}
%   \Description{Enjoying the baseball game from the third-base
%   seats. Ichiro Suzuki preparing to bat.}
%   \label{fig:teaser}
% \end{teaserfigure}

% \received{20 February 2007}
% \received[revised]{12 March 2009}
% \received[accepted]{5 June 2009}

%%
%% This command processes the author and affiliation and title
%% information and builds the first part of the formatted document.
\maketitle

\section{Introduction}

Large Language Models (LLMs) have recently demonstrated strong potential in recommender systems \cite{10.1145/3678004,zhao2024recommender,wu2024survey,wang2024towards}, attracting increasing research attention. Prior studies have integrated LLMs into various stages of the recommendation pipeline, particularly for enriching user and item representations. By leveraging their generative and semantic capabilities, LLMs can produce synthetic features or textual side information to enhance the modeling of user preferences and item characteristics \cite{10.1145/3701716.3735085,ren2024representation,liu2024modeling,liu2025understanding,liu2024once,gao2024generative}. Other works utilize LLM embedding spaces to initialize user and item representations \cite{10.1145/3404835.3463069,qiu2021u,doddapaneni2024user}, enabling recommender models to benefit from semantically informed representations.

Beyond feature augmentation, LLMs have also been explored as standalone recommenders that directly generate candidate items through prompting and reasoning. Prompt-based methods provide LLMs with user profiles and candidate item information, allowing them to select suitable items without task-specific fine-tuning. However, this paradigm suffers from long context length issues and information noise when large-scale user–item data are included, often leading to degraded recommendation performance. Alternatively, some studies fine-tune LLMs using supervised learning to align them with recommendation tasks. While effective, such approaches are computationally expensive and impractical in real-world scenarios due to the large model size and frequent updates required.

Despite these advances, collaborative filtering (CF), a fundamental signal in recommender systems, has not been explicitly or effectively modeled in most LLM-based approaches. LLMs struggle to capture implicit user–item interaction patterns solely through prompting or fine-tuning. To address this limitation, several studies bridge CF and LLMs by injecting collaborative knowledge from pretrained CF models  \cite{zhang2025collm} or collaborative evidence in the prompt \cite{CoRAL}. For example, ColaRec \cite{wang2024content} integrates item content and CF signals by generating item identifiers derived from pretrained CF models, while A-LLMRec \cite{kim2024large} transfers collaborative knowledge into LLMs to improve both cold-start and warm recommendation performance. However, these approaches essentially attach CF outputs to LLM text embeddings, heavily rely on pretrained CF models, limiting end-to-end learning and adaptability to dynamic user–item interactions. 

Recently, agentic LLMs have emerged as autonomous frameworks capable of decomposing complex tasks into multi-step reasoning processes, such as planning and decision-making. In recommender systems, agentic LLMs typically employ profile and memory modules to simulate user behavior and support personalized recommendations \cite{zhang2024generative,wang2024recmind}. While these memory-based agents are effective in capturing high-level user preferences, existing designs mainly store descriptive or semantic information and overlook collaborative filtering signals embedded in historical interactions.

To address this gap, we propose a novel agentic LLM-based recommender that leverages an evolving memory module to explicitly model collaborative filtering information. Our memory design is inspired by A-Mem \cite{xu2025mem}. Specifically, each user behavior is stored as a memory entry in a database. When a new, similar behavior arrives, the corresponding memory entry is updated and refined. Through this continual evolution process, the memory module aggregates recurring interaction patterns, enabling LLMs to capture implicit collaborative signals without relying on pretrained CF models.

Our main contributions are summarized as follows:
\begin{itemize}
    \item We propose a novel form of semantic collaborative filtering \textbf{AMEM4Rec}, where cross-user interaction patterns are distilled into textual memory representations. Unlike matrix factorization-based CF that learns from explicit co-occurrence matrices, our approach captures implicit collaborative signals through iterative semantic pattern aggregation across users.
    \item We design a memory evolution strategy that aggregates similar user behaviors over time, enabling LLMs to capture implicit collaborative patterns in an end-to-end manner. We design a dual validator memory evolution mechanism that combines similarity-based and semantic-based validation to selectively link and update cross-user memories. This enables the system to distill shared behavioral patterns while filtering out noise.
    \item Extensive experiments on four real-world datasets (Amazon Fashion, Video Games, 
CDs \& Vinyl, and MIND) demonstrate that AMEM4Rec consistently outperforms state-of-the-art LLM-based recommenders across all metrics, with particularly 
strong improvements in sparse interaction scenarios.
\end{itemize}
% \section{Related Work}
% \subsection{LLM Integration in Recommender Systems}
% \subsection{Hybrid Methods Bridging CF and LLMs}
% \subsection{Agentic LLM-based Recommender Systems}
\section{Related Work}

% We review prior work in three main directions: LLM-based recommender systems, collaborative filtering integration with LLMs, and agentic/memory-based approaches.
\subsection{LLM-based Recommender Systems}

Large Language Models (LLMs) have been increasingly integrated into recommender systems to leverage their strong semantic understanding and generative capabilities. Early approaches primarily use LLMs for feature augmentation, generating synthetic textual side information or enriching user/item representations \cite{zhao2024recommender,wu2024survey,ren2024representation,liu2024modeling,gao2024generative}. For example, methods such as U-BERT \cite{qiu2021u} and Chat-Rec \cite{gao2023chat} utilize LLM embeddings to initialize user and item representations, benefiting from semantically informed starting points. More recently, LLMs have been explored as standalone recommenders that directly generate or rank items through prompting. Prompt-based methods provide LLMs with user profiles and candidate information, allowing zero-shot or few-shot recommendation without task-specific fine-tuning \cite{10.1145/3678004,wang2024towards}. Generative approaches, such as P5 \cite{p5} and TALLRec \cite{bao2023tallrec}, fine-tune LLMs to produce item IDs or explanations. However, these methods often suffer from long context length limitations, information noise in large-scale inputs, and increased hallucination risks when handling extensive user–item data.

\subsection{Collaborative Filtering Integration with LLMs}

Collaborative filtering (CF) remains a core signal in traditional recommender systems, capturing implicit user–item co-occurrence patterns \cite{koren2009matrix,he2020lightgcn}. Integrating CF into LLM-based recommenders is challenging, as LLMs primarily excel at semantic reasoning and struggle to learn implicit interaction signals from sparse data without explicit guidance.

Recent works have explored the direction to bridge CF and LLMs. Specifically, it relies on injecting collaborative knowledge from pretrained CF models. ColaRec \cite{wang2024content} generates item identifiers conditioned on both content and CF signals derived from pretrained models (e.g., LightGCN or SASRec), while A-LLMRec \cite{kim2024large} transfers collaborative knowledge from pretrained recommenders into LLMs to improve cold-start and warm performance. These approaches achieve strong results but introduce dependency on external CF architectures, limiting end-to-end learning and adaptability to new domains or dynamic datasets.

Despite these advances, existing methods either depend heavily on pretrained CF models or fail to explicitly capture collaborative signals in an end-to-end manner, particularly in sparse interaction settings where semantic knowledge alone is insufficient.

\subsection{Agentic and Memory-based Recommender Systems}

Agentic LLMs, capable of multi-step reasoning and planning, have recently been applied to recommender systems to simulate user behavior and support personalized recommendations \cite{wang2024recmind,zhang2024generative}. These frameworks typically employ profile and memory modules to store high-level user preferences and historical context. Memory-augmented agents, including AgentCF~\cite{zhang2024agentcf}, attempt to capture collaborative signals and store them as user profiles in memory.

However, existing memory designs in agentic recommenders primarily focus on descriptive or semantic information and rarely capture collaborative filtering signals embedded in cross-user interaction patterns. To the best of our knowledge, no prior work has explicitly leveraged memory evolution to aggregate implicit collaborative signals in an end-to-end fashion without relying on pretrained CF models.

Our proposed AMEM4Rec addresses this gap by introducing a memory evolution mechanism tailored for cross-user collaborative pattern aggregation, enabling robust recommendation performance in sparse and dynamic scenarios.
\section{Methodology}
\begin{figure*}[!ht]
    \centering
    \includegraphics[width=0.9\linewidth]{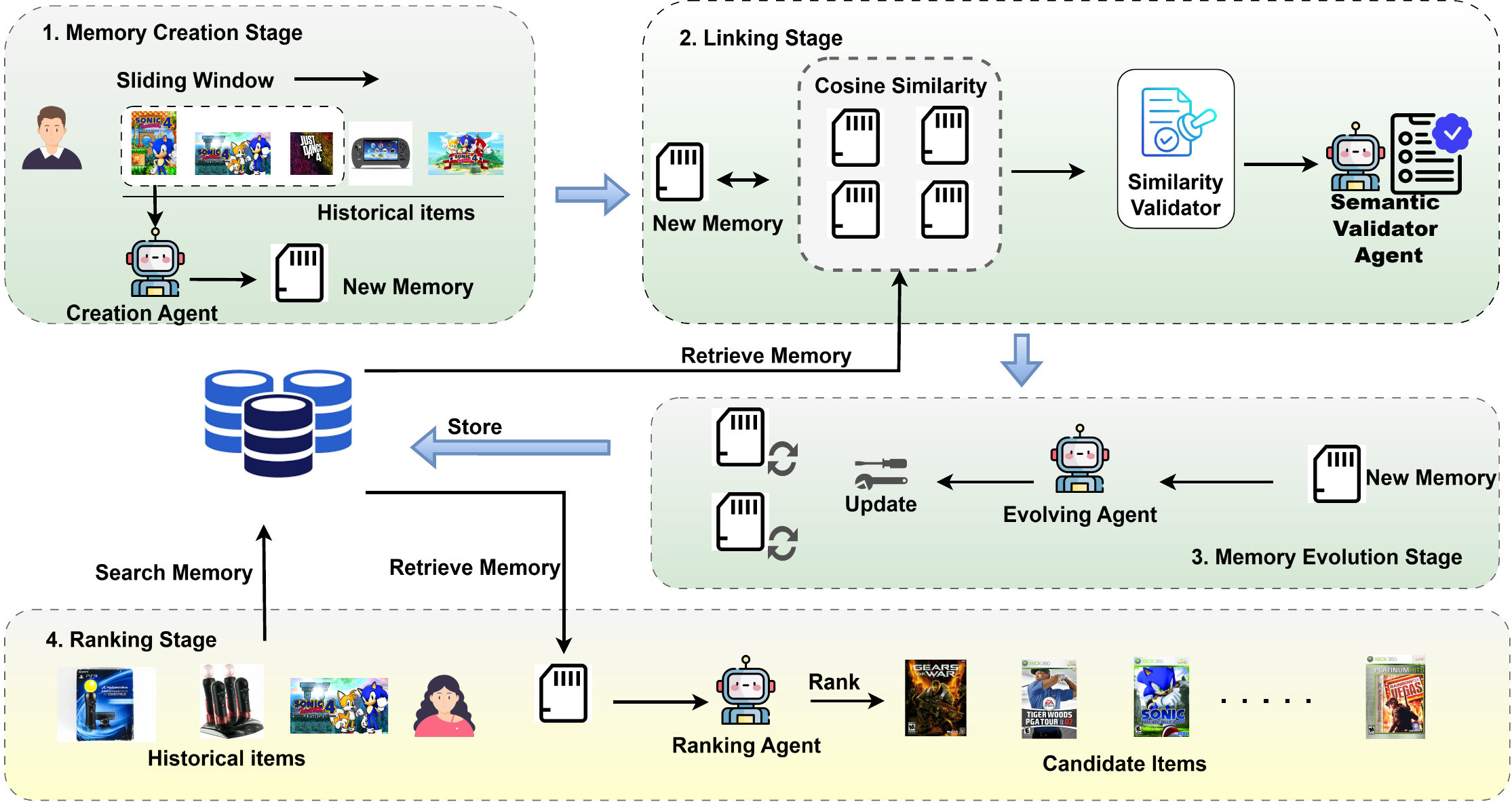}
    % \includesvg[width=0.9\linewidth]{figs/MemA_Rec.svg}
    \caption{Overview of the AMEM4Rec framework showing three training stages: (1) Memory 
Creation from user interaction histories, (2) Memory Linking via dual validators 
(similarity + semantic), and (3) Memory Evolution through iterative updates. 
During inference, retrieved memories augment the LLM agent for personalized ranking.}
    \label{fig:framework}
\end{figure*}
This section presents our proposed approach, AMEM4Rec, as shown in Figure \ref{fig:framework}. The framework consists of three stages for training the memory pool, followed by a final ranking stage during inference. Specifically, we describe the memory creation stage in Subsection \ref{subsec:memory_creation}, the linking and validation stage in Subsection \ref{subsec:linkmemory}, and the memory evolution stage in Subsection \ref{subsec:memory_evo}. During inference, the trained memory system is used for ranking, as detailed in Subsection \ref{subsection:ranking}.

In this section, we explain how memory augmentation supports agentic recommendation. Collaborative filtering (CF) signals are embedded into the memory pool to enhance the agent's autonomous reasoning during the re-ranking task. Unlike AgentCF, which maintains isolated memories per user and often misses shared collaborative patterns, we design a unified memory system that encodes latent, abstract group-level behaviors. This enables the agent to better understand and generalize user patterns across the population. The knowledge stored in the memory pool represents distilled behavioral patterns that enrich the agent's multi-step decision-making process. To capture collaborative signals, we extract insights from users' historical interactions using a sliding window approach. This process generates a large number of memory fragments from multiple users, which are continuously updated and refined. Ultimately, each memory represents a shared behavioral pattern of a user group, serving as a foundation for informed and collaborative-aware recommendations.

% In this section, we 
\subsection{Preliminaries}
\subsubsection{Problem Formulation}
In this work, we introduce a re-ranking framework that incorporates memory of user behavior to refine and rank the candidate items more effectively. To set the stage, we first briefly review the conventional recommendation task before detailing our proposed ranking approach.

In a typical recommender system, the user set is denoted as $\mathcal{U} = \{u_1, u_2, \dots, u_n\}$, the item set as $\mathcal{I} = \{i_1, i_2, \dots, i_m\}$, and the observed interactions as $\mathcal{D} = \{(u, i)\}$, where each tuple represents a user-item interaction (e.g., view, click, purchase, or rating). The core goal of a recommendation model is to learn an approximate scoring function $f(u, i)$ that estimates the likelihood of user $u$ interacting with item $i$. Higher scores indicate stronger predicted preference, guiding the selection of top-ranked items for recommendation.

In our problem, we focus on a re-ranking recommendation task, where a set of candidate items is first retrieved by a base recommender (e.g., popularity-based), and the goal is to refine and reorder these candidates to produce a more accurate personalized ranking. Formally, given a user $u$ and a candidate set $\mathcal{C}_u = \{i_1, i_2, \dots, i_m\}$ of size $m$. Unlike traditional recommendation models that predict scores from scratch, our approach leverages the user's interaction history $h_u$ and the evolving memory pool $\mathcal{M}$ to augment the LLM agent's reasoning process. Specifically, we prompt the LLM to rank the candidates based on historical items and retrieved collaborative memories, thereby enhancing recommendation accuracy without requiring end-to-end retraining of a large model. This re-ranking paradigm is particularly effective in sparse and dynamic scenarios, where semantic understanding and cross-user collaborative signals play a critical role in overcoming the limitations of conventional scoring functions.
\begin{figure*}[t]
    \centering
    \includegraphics[width=0.7\linewidth]{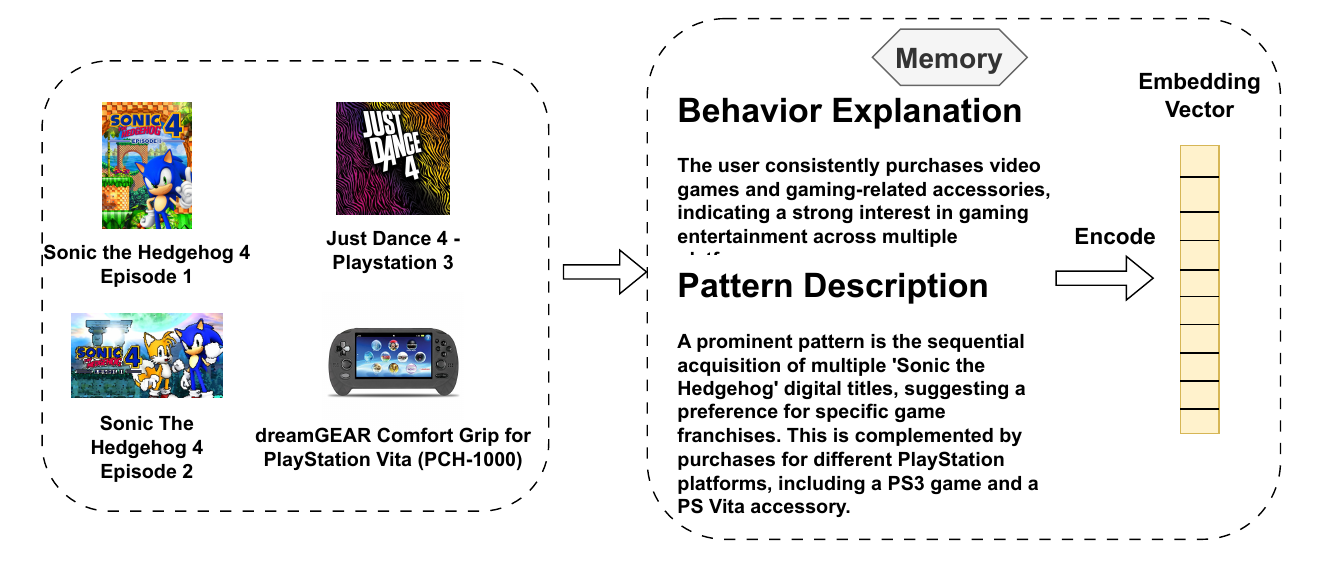}
    \caption{Illustration of extracted memory components: Behavior Explanation and Pattern Description.}
    \label{fig:memory_component}
\end{figure*}
\subsubsection{Collaborative Filtering}

Collaborative filtering (CF) has long been the cornerstone of traditional recommender systems, capturing implicit user–item interaction patterns through matrix factorization \cite{koren2009matrix} or sequential recommendation \cite{kang2018self}. However, integrating CF signals into LLM-based recommenders remains a significant challenge, as LLMs primarily excel at semantic understanding and struggle to learn implicit interaction patterns from sparse data without explicit guidance.

Despite these advances, existing approaches either depend heavily on pretrained CF models or fail to explicitly model collaborative signals in an end-to-end manner. Moreover, they rarely address the challenge of capturing implicit cross-user patterns in sparse datasets, where semantic knowledge alone is insufficient. In contrast, our proposed AMEM4Rec learns collaborative signals implicitly through cross-user memory evolution, eliminating the need for any pretrained CF model and enabling robust performance in sparse and dynamic scenarios.
\subsubsection{Learning Space in Agentic Recommendation}
In our approach, we do not fine-tune the LLM agent parameter as in neural network training. We distill a memory database, which contains semantic text. This text is trained over time when a new user arrives. More generally, following a previous work \cite{liu2025advances}, we present the parameter of the LLM agent as the mental state, as follows:

\begin{equation}
    M = \{M^{\theta}, M^{mem}\}
\end{equation}

In this combination, $M^{\theta}$ represents the core model LLM. This parameter contain the holistic knowledge about semantic information. Additionally, $M^{mem}$ represents the memory model. It is trained on the recommendation dataset to understand user behaviour. The word 'train' implies that the memory will be updated to catch the changes of the user. This component is side information to help the LLM agent make better decisions. 

% \subsubsection{Memory-Guided Collaborative Filtering}
% In this part, we present our explanation of using memory for agentic recommendation. Collaborative filtering information is integrated into memory to help the agent making a decision on the re-ranking task. Unlike AgentCF, which constructs a separate memory for each individual user, we observe that certain collaborative signals are not captured. We therefore aim to design a memory system that encodes latent and abstract group-level behaviors, enabling a better understanding of user patterns. The knowledge stored in the memory pool represents behavioral patterns that enrich the agent’s reasoning during recommendation decisions. Therefore, we aim to distill a shared knowledge pool that serves as a foundation for informed recommendation. To aware collaborative signals, we extract insights from users’ historical interactions using a sliding window. This process enables the agent to identify patterns from past behaviors and leverage them for future decisions. As a result, a large number of memory fragments are created and continuously updated from multiple users. We expect each memory to represent a shared behavioral pattern of a user group. 

% \begin{figure*}
%     \centering
%     \includegraphics[width=0.7\linewidth]{figs/Evolving_memory.pdf}
%     \caption{Caption}
%     \label{fig:placeholder}
% \end{figure*}
\subsection{Memory Definition}
\label{subsec:memory_creation}
In our approach, each memory fragment is constructed as a user behavior pattern. We apply a sliding window over the user's interaction sequence and extract patterns from each windowed cluster, which we assume capture insightful behaviors that improve the accuracy of predicting the next item. For each user $u$, we define their interaction history as a temporally ordered sequence 
$h_u = (i_{u,1}, i_{u,2}, \dots, i_{u,|h_u|})$, sorted by timestamp, where $i_{u,t}$ denotes the item interacted with by user $u$ at time step $t$. For each item $i_{u,t}$, we utilize its title $t_{i_{u,t}}$ and category $c_{i_{u,t}}$ as descriptive features. All generated behavior memories are stored in a shared global memory pool. 
We apply a sliding window of size $w$ over the sequence $h_u$. For the $k$-th window starting at position $t_k$, we feed the sequence of (title, category) pairs 
$\{(t_{i_{u,t_k}}, c_{i_{u,t_k}}), \dots, (t_{i_{u,t_k+w-1}}, c_{i_{u,t_k+w-1}})\}$ into the LLM agent via a carefully designed prompt. The LLM outputs a structured pattern description $p_k$, which consists of two textual components: a behavior explanation and a pattern description:

\begin{equation}
    p_k \leftarrow \text{Prompt}_{\text{LLM}} \bigl( (t_{i_{u,t_k}}, c_{i_{u,t_k}}), \dots, (t_{i_{u,t_k+w-1}}, c_{i_{u,t_k+w-1}}) \bigr)
\end{equation}

In Figure \ref{fig:memory_component}, we characterize the extracted memory knowledge using two complementary components. The behavior explanation captures the underlying user behavior by providing a concise description of the motivations behind item purchases. In contrast, the pattern representation summarizes recurring interaction structures shared across users, highlighting common preference trends and behavioral regularities. Together, these components enable the memory to encode both interpretable user intent and abstract collaborative patterns.

With the generated textual knowledge, we encode it into a shared embedding space to support subsequent retrieval modules. 
\begin{equation}
    e_k = \text{Encode}(p_k)
\end{equation}

For memory creation, we iterate this process over all users. We define each memory $m_k =(p_k, e_k)$. The memory pool is constructed as $M^{mem} = \{m_k\}^n$

% \subsection{Cross-User Linking and Evolution}
\subsection{Link Memory}
\label{subsec:linkmemory}
Following the creation of each new memory, we perform a linking operation. This step identifies similar memories in the global pool and triggers their evolution. The purpose of this evolution is to make existing memories more dynamic by incorporating insights from the new memory, thereby enabling each memory to reflect both historical patterns and emerging trends. Through linking, memories implicitly share information across users, allowing individual behaviors to propagate within the global memory pool. As a result, collaborative signals are gradually reinforced through the iterative evolution process. The linking process is carried out using two main components: a retrieval module and an agent module. When a new memory is created, we first employ its embedding to identify a set of similar memories in the global pool. After selecting the top-$  k  $ most related memories, their textual patterns are provided as input to the agent, which decides whether linking between the new memory and any of the retrieved ones is appropriate. 

Upon the addition of a new memory $  m_n  $ to the pool, we perform similarity search using cosine similarity. The cosine similarity is calculated between the new memory's embedding $  e_n  $ and every embedding $e_j$ in the current memory pool. The top-$  k  $ most similar memories are then selected as candidates.
\begin{equation}
    score_{n,j} = \frac{e_n\cdot e_j}{|e_n|\cdot |e_j|}
\end{equation}
After scoring, the system select top -$k$ memory from its rank:
\begin{equation}
    \mathcal{M}^n_{k-nn} = \{m_j|\text{rank}(\text{score}_{n,j}) \leq k , m_j \in M^{mem}\}
\end{equation}

Given the top-$k$ most similar memories $\mathcal{M}^n_{k\text{-nn}} = \{m_1, m_2, \ldots, m_k\}$ retrieved for a new memory $m^n$, we validate their similarity scores $\mathcal{S}_n = \{\text{score}_{n,1}, \text{score}_{n,2}, \ldots, \text{score}_{n,k}\}$ to determine the optimal memory management strategy. Rather than using a single hard threshold, we introduce a \textbf{soft threshold mechanism} for the similarity validator, incorporating multiple decision strategies to enable more nuanced control over memory updates and storage. 

We define two threshold boundaries, $\tau_{\text{low}}$ and $\tau_{\text{high}}$ (where $\tau_{\text{low}} < \tau_{\text{high}}$), which partition the similarity space into three distinct operational zones. Let $s_{\max} = \max(\mathcal{S}_n)$ denote the maximum similarity score. We further analyze the distribution of similarity scores by computing:
\begin{equation}
\begin{aligned}
p_{\text{high}} &= \frac{|\{s \in \mathcal{S}_n : s \ge \tau_{\text{high}}\}|}{k}, \\
p_{\text{medium}} &= \frac{|\{s \in \mathcal{S}_n : \tau_{\text{low}} \le s < \tau_{\text{high}}\}|}{k}, \\
p_{\text{low}} &= \frac{|\{s \in \mathcal{S}_n : s < \tau_{\text{low}}\}|}{k},
\end{aligned}
\end{equation}
representing the proportions of high, medium, and low similarity scores, respectively. The memory management strategy $\pi(m_n, \mathcal{M}^n_{k\text{-nn}})$ is determined by the following decision tree:
\begin{equation}
\small
\begin{aligned}
\pi(m_n, \mathcal{M}^n_{k\text{-nn}})
=
\begin{cases}
\text{STORE}_{\text{only}}, & \text{if } s_{\max} < \tau_{\text{low}}, \\[4pt]
\text{UPDATE}_{\text{and STORE}}, & \text{if } \tau_{\text{low}} \le s_{\max} < \tau_{\text{high}}, \\[4pt]
\text{UPDATE}_{\text{only}}, & \text{if } s_{\max} \ge \tau_{\text{high}} \text{ and } p_{\text{high}} \ge 0.6, \\[4pt]
\text{STORE}_{\text{only}}, & \text{if } s_{\max} \ge \tau_{\text{high}} \text{ and } p_{\text{low}} \ge 0.5, \\[4pt]
\text{UPDATE}_{\text{and STORE}}, & \text{otherwise}.
\end{cases}
\end{aligned}
\end{equation}

This strategy aims to preserve the generality and representativeness of the memory system through distribution-aware decision making.
When $s_{\max} < \tau_{\text{low}}$, the new memory is sufficiently dissimilar and is stored without updates to prevent introducing false knowledge.
When $s_{\max} \ge \tau_{\text{high}}$ with a high concentration of similar memories ($p_{\text{high}} \ge 0.6$), we only update existing memories to avoid redundancy.
Conversely, when $p_{\text{low}} \ge 0.5$ despite high $s_{\max}$, most retrieved memories are actually dissimilar, suggesting the new memory represents a distinct pattern worthy of storage.
In intermediate cases, we update the memory with scores above the low threshold and store new memories, capturing novel information while maintaining connections to existing knowledge. This approach enriches the memory pool with informative insights while managing storage costs.

After applying the similarity validator to the set of top-$k$ nearest neighbors $\mathcal{M}^n_{k\text{-nn}}$, we next introduce the semantic validator agent, which evaluates which memories should be updated based on the new memory. Notably, this decision is made by analyzing the semantic content of the textual patterns, rather than relying solely on cosine similarity. 
\begin{equation}
    \mathcal{L}=\{m_i,..., m_l\} \leftarrow \text{Prompt}_{LLM}(m_n, \mathcal{M}^n_{k\text{-nn}}| \pi(m_n,\mathcal{M}^n_{k\text{-nn}})), l\leq k
\end{equation}
Here, $ \mathcal{L}$ is a set of memories that are linked to the new memory. 
\subsection{Memory Evolution}
\label{subsec:memory_evo}

\begin{figure}[t]
    \centering
    \begin{tabular}{cc}
        \includegraphics[width=1\linewidth]{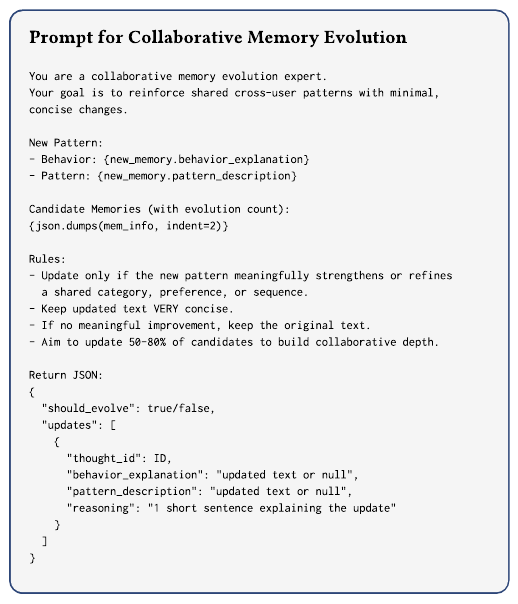} 
    \end{tabular}
    \caption{Prompt template for updating memory}
    \label{fig:promptUpdate}
\end{figure}

The link generation step identifies the most relevant existing memories as candidates for the evolution process. These candidates are collected in the set $\mathcal{L}$. The final decision on which memory should be updated is made by a separate agent. We prompt the agent to select the memory and update it as follows:
\begin{equation}
    m^*_i \leftarrow \text{Prompt}_{\text{LLM}}(m_n, m_i), \quad m_i \in \mathcal{L}
\end{equation}
Here, $m^*$ denotes the updated memory, which directly replaces $m_i$ in the memory pool. Figure~\ref{fig:promptUpdate} illustrates the prompt used for memory updating.
% \begin{table*}[ht]
% \centering

% \begin{tabular}{lccccccccccccc}
% \hline
%  & \multicolumn{3}{c}{\textbf{Fashion}} & \multicolumn{3}{c}{\textbf{CDs and Vinyl}} & \multicolumn{3}{c}{\textbf{Video Game}} & \multicolumn{3}{c}{\textbf{MIND}} \\
% \cmidrule(lr){2-4} \cmidrule(lr){5-7} \cmidrule(lr){8-10} \cmidrule(lr){11-13}
% Method & N@1 & N@5 & N@10 & N@1 & N@5 & N@10 & N@1 & N@5 & N@10 & N@1 & N@5 & N@10 \\
% \hline

% LLMRank   &0.1700 &0.2902 &0.4027 &0.1700 &0.3703 & 0.4296 & 0.3700& 0.5748 & 0.6095 & 0.1000&0.2696 &0.3304 \\
% AgentCF &0.1700 & 0.3029 &0.3735 & 0.1500 & 0.3318 & 0.4072 &0.1100 &0.2972 & 0.3679& 0.1211&0.3065 &0.3490 \\
% iAgent &0.1400 &0.2716 & 0.3337& 0.1900&0.3553 & 0.4192 &0.2800 &0.4237 &0.4809 & 0.1383& 0.2905 &0.3257 \\
% \hline
% % AMEM4Rec &  0.3000 & 0.4520 &0.5118 &0.2400 &0.4116 &0.4638 &0.3800 &0.5561 &0.5861  &0.1474 &0.3441 &0.3892 \\
% AMEM4Rec  &  0.1800 & 0.3092 &0.4038 &0.2400 &0.4116 &0.4638 &0.3900 &0.5754 &0.6108  &0.1789 & 0.3607 &0.4150 \\
% % Method9 & & & & & & & & & & & & \\
% % Method10 & & & & & & & & & & & & \\
% \hline
% \end{tabular}
% \caption{Performance comparison across different datasets and metrics}
% \label{tab:results}
% \end{table*}
\begin{table*}[ht]
\centering
\small
\setlength{\tabcolsep}{6pt}
\begin{tabular}{lcccccccccccc}
\toprule
 & \multicolumn{3}{c}{\textbf{Fashion}} 
 & \multicolumn{3}{c}{\textbf{CDs and Vinyl}} 
 & \multicolumn{3}{c}{\textbf{Video Game}} 
 & \multicolumn{3}{c}{\textbf{MIND}} \\
\cmidrule(lr){2-4} \cmidrule(lr){5-7} \cmidrule(lr){8-10} \cmidrule(lr){11-13}
\textbf{Method} 
& N@1$\uparrow$ & N@5$\uparrow$ & N@10$\uparrow$
& N@1$\uparrow$ & N@5$\uparrow$ & N@10$\uparrow$
& N@1$\uparrow$ & N@5$\uparrow$ & N@10$\uparrow$
& N@1$\uparrow$ & N@5$\uparrow$ & N@10$\uparrow$ \\
\midrule
LLMRank   
& 0.1700 & 0.2902 & 0.4027 
& 0.1700 & 0.3703 & 0.4296 
& 0.3700 & 0.5748 & 0.6095 
& 0.1000 & 0.2696 & 0.3304 \\

AgentCF 
& 0.1700 & 0.3029 & 0.3735 
& 0.1500 & 0.3318 & 0.4072 
& 0.1100 & 0.2972 & 0.3679 
& 0.1211 & 0.3065 & 0.3490 \\

iAgent 
& 0.1400 & 0.2716 & 0.3337 
& 0.1900 & 0.3553 & 0.4192 
& 0.2800 & 0.4237 & 0.4809 
& 0.1383 & 0.2905 & 0.3257 \\

\midrule
\textbf{AMEM4Rec}  
& \textbf{0.1800} & \textbf{0.3092} & \textbf{0.4038}
& \textbf{0.2400} & \textbf{0.4116} & \textbf{0.4638}
& \textbf{0.3900} & \textbf{0.5754} & \textbf{0.6108}
& \textbf{0.1789} & \textbf{0.3607} & \textbf{0.4150} \\
\bottomrule
\end{tabular}
\caption{Overall performance comparison on ranking recommendation task. AMEM4Rec 
consistently outperforms baselines across all datasets and metrics 
($NDCG@K, K \in {1,5,10}$). Higher values indicate better performance.}
\label{tab:results}
\end{table*}

\subsection{Memory-Augmented Ranking}
\label{subsection:ranking}
For the ranking recommendation task, we prompt the agent to reorder the candidate set and identify the most suitable next item. The input includes the user’s historical items, relevant memory fragments, and the candidate list. To retrieve pertinent memories, we encode the user’s recent history (titles and categories) into the embedding space, then compute cosine similarity scores to select the most relevant memories for the ranking process. 
We implement AMEM4Rec following Algorithm~\ref{alg:amem4rec}, which summarizes the overall procedures of the proposed framework.

\section{Experiments}
\begin{algorithm}[t]
\caption{AMEM4Rec Framework}
\label{alg:amem4rec}
\begin{algorithmic}[1]
\REQUIRE Users $\mathcal{U}$, window size $w$, thresholds $\tau_{\text{low}}$, $\tau_{\text{high}}$, top-$k$
\ENSURE Ranked recommendations $\mathcal{R}_u$
\STATE \textbf{Training:} $\mathcal{M}_{\text{mem}} \leftarrow \emptyset$
\FOR{each user $u \in \mathcal{U}$}
    \FOR{each window $W_t$ of size $w$ in history $h_u$}
        \STATE $m_t \leftarrow (\text{Prompt}_{\text{LLM}}(W_t), \text{Encode}(\cdot))$ \COMMENT{Create memory}
        \STATE $\mathcal{M}_{k\text{-nn}} \leftarrow \text{Top-}k\text{ from } \mathcal{M}_{\text{mem}} \text{ by cosine similarity}$
        \STATE $\mathcal{S}_t \leftarrow \{\text{score}_{t,1}, \ldots, \text{score}_{t,k}\}$ \COMMENT{Similarity scores}
        \STATE $s_{\max} \leftarrow \max(\mathcal{S}_t)$
        \STATE Compute $p_{\text{high}}, p_{\text{medium}}, p_{\text{low}}$ from $\mathcal{S}_t$ \COMMENT{Score distribution}
        \STATE $(\text{doUpdate}, \text{doStore}) \leftarrow \pi(m_n, \mathcal{M}^n_{k\text{-nn}})$
        \IF{$\text{doUpdate}$}
            \STATE $\mathcal{L} \leftarrow \text{Prompt}_{\text{LLM}}(m_t, \mathcal{M}_{k\text{-nn}})$ \COMMENT{Validate links}
            \FOR{$m_i \in \mathcal{L}$}
                \STATE $m_i^* \leftarrow \text{Prompt}_{\text{LLM}}(m_t, m_i)$ \COMMENT{Evolve memory}
                \STATE $\mathcal{M}_{\text{mem}} \leftarrow \mathcal{M}_{\text{mem}} \setminus \{m_i\} \cup \{m_i^*\}$
            \ENDFOR
        \ENDIF
        \IF{$\text{doStore}$}
            \STATE $\mathcal{M}_{\text{mem}} \leftarrow \mathcal{M}_{\text{mem}} \cup \{m_t\}$ \COMMENT{Store new memory}
        \ENDIF
    \ENDFOR
\ENDFOR
\STATE \textbf{Inference:}
\FOR{each user $u \in \mathcal{U}$}
    \STATE $\mathcal{M}_{\text{retrieved}} \leftarrow \text{Top-}k_{\text{mem}}\text{ memories by similarity to } h_u$
    \STATE $\mathcal{R}_u \leftarrow \text{Prompt}_{\text{LLM}}(h_u, \mathcal{M}_{\text{retrieved}}, \mathcal{C}_u)$ \COMMENT{Rank candidates}
\ENDFOR
\RETURN $\{\mathcal{R}_u\}_{u \in \mathcal{U}}$
\end{algorithmic}
\end{algorithm}

\subsection{Experimental Setup}
\subsubsection{Datatasets}
To evaluate the effectiveness of our approach, we conduct extensive experiments on four datasets: Fashion, Video Games, CDs and Vinyl (Amazon review dataset~\cite{ni2019justifying}), and MIND (news dataset~\cite{wu2020mind}). Due to the high cost of API calls, we randomly sample 300 users for both memory training and evaluation. We choose a set of users who interact with more than 10 items. For data preparation, we adopt the leave-one-out strategy, where the last item in each user's interaction history is reserved as the test item, while the remaining items are used to train the memory system. We construct a candidate set of 20 items for evaluation by shuffling the ground-truth test item with negatively sampled items from the dataset.
\subsubsection{Hyperparameter Setup}
We employ the Gemini-2.5-Flash API as the LLM backbone for all experiments. The window size is set to $w = 3$, and the linking stage retrieves the top-$5$ memory entries. The similarity thresholds are fixed at $\tau_{\text{low}} = 0.55$ and $\tau_{\text{high}} = 0.9$. We employ a Sentence-BERT encoder \cite{reimers2019sentence} to generate embeddings for textual memories. The prompts used in all experiments are provided in Appendix~\ref{apendix:prompt}.

\subsubsection{Evaluation Metrics}
To evaluate the performance, we employ NDCG@K (K=1,5,10) as a metric to measure the effectiveness of the ranking task.
\subsubsection{Baseline Models}
We compare our method with several baseline approaches, including LLMRank \cite{hou2024large}, which is the first study to employ large language models as zero-shot ranking recommenders.AgentCF \cite{zhang2024agentcf} is an agent-based recommender that maintains user profiles in agentic memory, learned via contrastive learning on semantic text. iAgent \cite{xu-etal-2025-iagent} proposes a framework that incorporates internal and external knowledge instructions into ranking prompts, supported by a memory module.
% \begin{itemize}
%     \item Training-free methods: BPR-MF \cite{rendle2012bpr}, SASRec \cite{kang2018self}
%     \item LLMs and Agentic Recommendation: LLMRank \cite{hou2024large}, AgentCF \cite{zhang2024agentcf}, iAgent \cite{xu-etal-2025-iagent}.
% \end{itemize}

\subsection{Overall Performance}
In table ~\ref{tab:results}, Our AMEM4Rec consistently outperforms all baseline methods across all scenarios, demonstrating its effectiveness for ranking-based recommendation. The constructed memory pool provides rich and informative knowledge that enhances the recommendation process. Through cross-user training, the memory module is able to capture collaborative signals, which in turn helps the model produce more accurate predictions.

% \begin{table*}[t]
% \centering

% \begin{tabular}{lccccccccccccc}
% \hline
%  & \multicolumn{3}{c}{Books -> CDs and Vinyl} & \multicolumn{3}{c}{Books -> Video Games} & \multicolumn{3}{c}{CDs and Vinyl -> Video Game} & \multicolumn{3}{c}{CDs and Vinyl -> Books} \\
% \cmidrule(lr){2-4} \cmidrule(lr){5-7} \cmidrule(lr){8-10} \cmidrule(lr){11-13}
% Method & N@1 & N@5 & N@10 & N@1 & N@5 & N@10 & N@1 & N@5 & N@10 & N@1 & N@5 & N@10 \\
% \hline

% LLMRank   & 0.3000 & 0.5281 & 0.5748&0.3600 & 0.5114 & 0.5636&0.3200 &0.5214&0.5584 &0.1000 &0.2696 & 0.3304 \\
% AgentCF & & & & & & & & & & & & \\
% iAgent & & & & & & & & & & & & \\
% \hline
% AMEM4Rec &  0.3000 & 0.4520 &0.5118 &0.2400 &0.4116 &0.4638 &0.3800 &0.5561 &0.5861  &0.1474 &0.3441 &0.3892 \\
% % Method9 & & & & & & & & & & & & \\
% % Method10 & & & & & & & & & & & & \\
% \hline
% \end{tabular}
% \caption{Cross domain}
% \label{tab:results2}
% \end{table*}

% \subsection{Further Analyses}
\subsection{Memory Construction Performance}
To demonstrate the effectiveness of our memory system for ranking tasks, we conduct experiments comparing full memory linking and evolution against a baseline without link generation and evolution. In the configuration without memory evolution, memories are created only from historical items and are not updated or evolved over time. This setting prevents any inter-memory connections, resulting in isolated, standalone memories where no collaborative signals are formed. In the setting without the semantic validator, the agent does not perform semantic reflection validation. Similarly, removing the similarity validator eliminates the use of the cosine similarity strategy. All memories are retrieved solely based on cosine similarity and then passed into the evolving process.

As shown in Table \ref{tab:ablation}, our model consistently outperforms both the variants. The complete system achieves better performance by incorporating both the similarity validator and the semantic validator and memory evolution. At each stage, mechanisms are applied to validate whether memories should be linked or updated over time. This result shows that each component is essential for identifying and extracting truly important knowledge.
\begin{figure*}[!ht]
    \centering
    \begin{tabular}{cc}
        \includegraphics[width=0.43\linewidth]{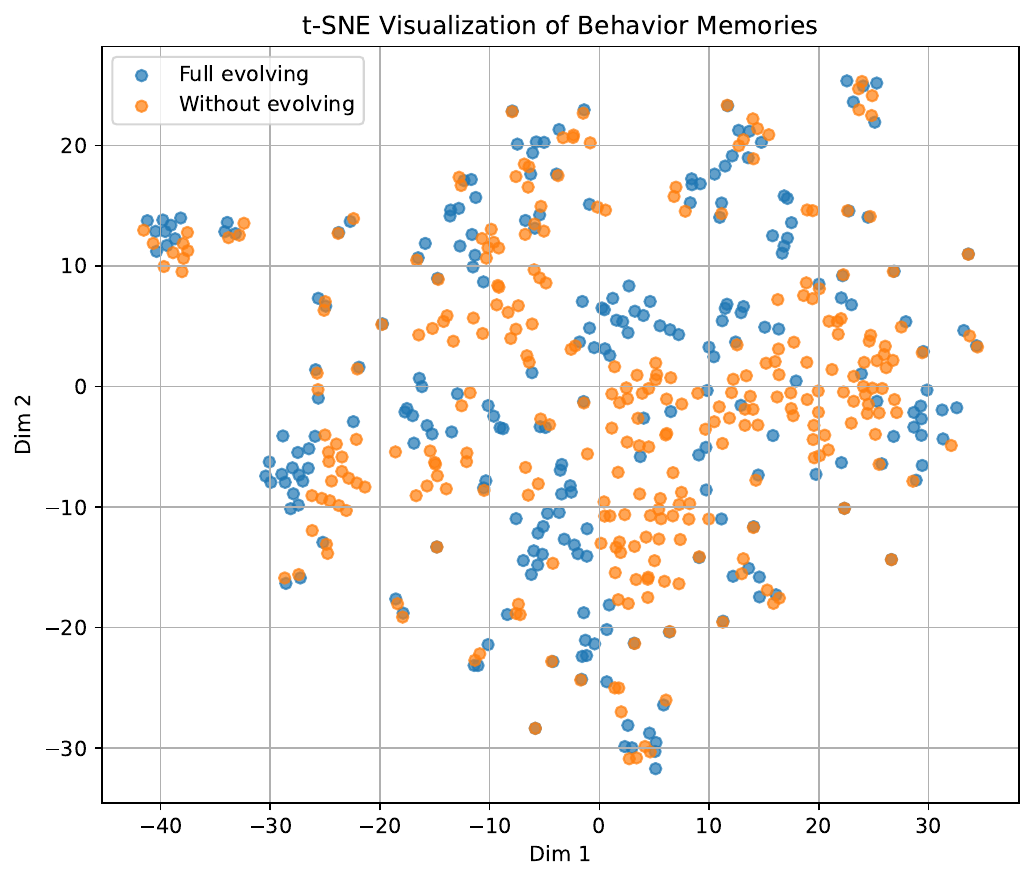} &
        \includegraphics[width=0.43\linewidth]{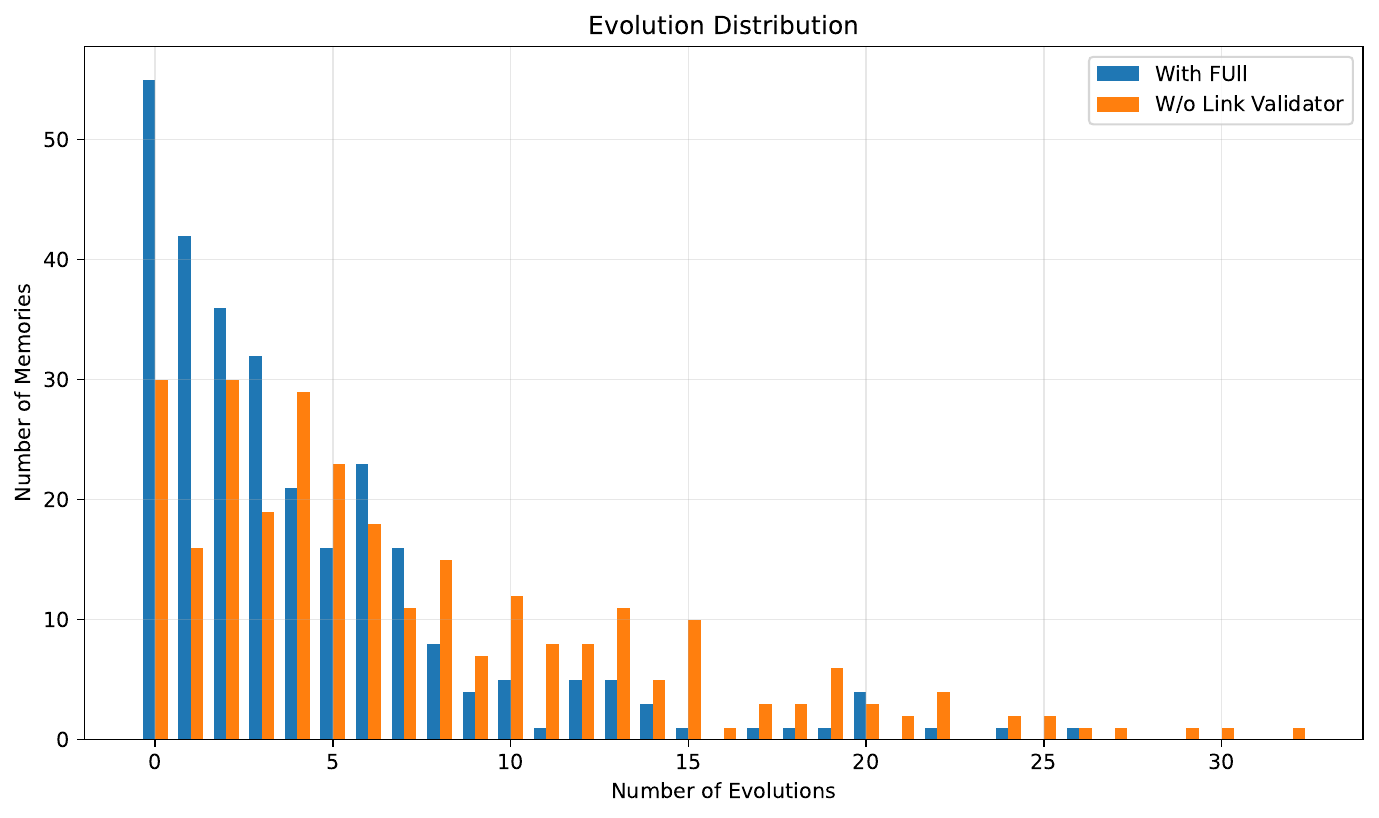} \\
        \small (a) t-SNE: Full evolution vs. without evolving & \small (b) Evolution count distribution: With vs. without Link Generation
    \end{tabular}
    \caption{Ablation study visualizations: (a) t-SNE embedding comparison between models with and without memory evolution mechanism, and (b) distribution comparison between full model and variant without the semantic validator.}
    \label{fig:tsne_compare}
\end{figure*}
\begin{table*}[t]
\centering
\small
\setlength{\tabcolsep}{6pt}
\begin{tabular}{lcccccc}
\toprule
\textbf{Method}
& \multicolumn{3}{c}{\textbf{Video Games}}
& \multicolumn{3}{c}{\textbf{MIND}} \\
\cmidrule(lr){2-4} \cmidrule(lr){5-7}
& NDCG@1$\uparrow$ & NDCG@5$\uparrow$ & NDCG@10$\uparrow$
& NDCG@1$\uparrow$ & NDCG@5$\uparrow$ & NDCG@10$\uparrow$ \\
\midrule
w/o Similarity Validator
& 0.3300 & 0.5136 & 0.5694
& 0.1800 & 0.3506 & 0.4110 \\

w/o Semantic Validator
& 0.3500 & 0.5163 & 0.5736
& 0.1500 & 0.3247 & 0.3957 \\

w/o Memory Evolution
& 0.3500 & 0.5260 & 0.5801
& 0.1474 & 0.3441 & 0.3892 \\
\midrule
\textbf{Full AMEM4Rec}
& \textbf{0.3900} & \textbf{0.5754} & \textbf{0.6108}
& \textbf{0.1900} & \textbf{0.3528} & \textbf{0.4245} \\
\bottomrule
\end{tabular}
\caption{Ablation study on the impact of similarity validation, semantic validation, and memory evolving. Higher NDCG values indicate better performance.}
\label{tab:ablation}
\end{table*}

% \begin{table*}[t]
% \centering
% \begin{tabular}{lcccccc}
% \toprule
% \bfseries Method
% & \multicolumn{3}{c}{\bfseries Video Games}
% & \multicolumn{3}{c}{\bfseries MIND} \\
% \cmidrule(lr){2-4} \cmidrule(lr){5-7}
% & NDCG@1 & NDCG@5 & NDCG@10
% & NDCG@1 & NDCG@5 & NDCG@10 \\
% \midrule
% w/o Similarity Validator
% & 0.3300 & 0.5136 & 0.5694
% & 0.1800 & 0.3506 & 0.4110\\
% w/o Semantic Validator
% & 0.3500 & 0.5163& 0.5736
% & 0.1500 & 0.3247 & 0.3957\\
% w/o Memory Evolving
% & 0.3500 & 0.5260 & 0.5801
% & 0.1474 & 0.3441 & 0.3892 \\

% \bfseries Full AMEM4Rec
% & \bfseries0.3900 & \bfseries 0.5754 & \bfseries 0.6108
% & \bfseries 0.1900 & \bfseries 0.3528 & \bfseries 0.4245 \\
% \bottomrule
% \end{tabular}
% \caption{Ablation study on the impact of memory evolving and link generation. Bold values indicate the best performance.}
% \label{tab:ablation}
% \end{table*}

(1) w/o Memory Evolution: Interestingly, the w/o Memory Evolution variant achieves performance quite close to the full model. This indicates that the initial memory creation process is already effective in capturing useful behavioral patterns for recommendation. However, the consistent gains of the full model demonstrate that iterative evolution further strengthens cross-user collaborative signals, enabling more robust generalization in sparse and dynamic scenarios. Moreover, Figure~\ref{fig:tsne_compare}(a) presents the t-SNE visualization of memory embeddings. In the full evolving model (blue points), memories exhibit clearer clustering and grouping compared to the without-evolving baseline (orange points). This indicates that the iterative evolution process successfully reinforces similar behavioral patterns across users, pulling related memories into closer regions in the embedding space. These observations confirm that memory evolution plays a crucial role in aggregating cross-user collaborative signals, enabling the memory pool to develop more structured and generalizable representations.

(2) w/o Semantic Validator: In Figure~\ref{fig:tsne_compare}(b), we present the distribution of evolution counts for the full model and the variant without the semantic validator. This variant exhibits a distribution skewed toward higher evolution counts, indicating that many memories undergo frequent updates. However, despite this increased number of evolutions, its overall performance remains inferior to the full model. We attribute this to the absence of validation from the link generation agent. Without the semantic filtering provided by linking, irrelevant or noisy memories are more likely to be included in the evolution process. As a result, unrelated patterns are propagated and reinforced, introducing noise into the memory pool and degrading the quality of collaborative signals. This observation highlights the critical role of the link generation step in ensuring that only meaningful cross-user patterns are evolved, thereby preserving the integrity and effectiveness of the memory pool.

(3) w/o Similarity Validator: Consistently, removing the intelligence strategy from AMEM4Rec leads to degraded performance. Without this mechanism, noisy or incorrect information can be propagated to the semantic validator. As a result, the semantic validator may introduce additional diversity into the memory based on unreliable signals. Although this increases memory diversity, it ultimately harms recommendation performance. 

\subsection{Further Analyses}
\begin{figure*}[!ht]
    \centering
    \includegraphics[width=0.7\linewidth]{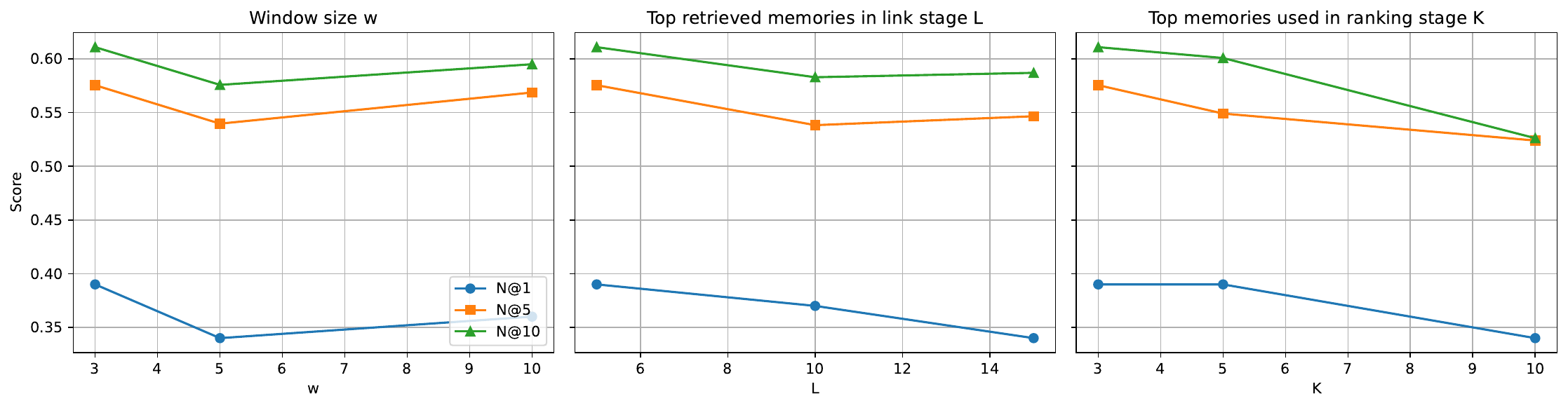}
    \caption{Hyperparameter sensitivity analysis on Video Games dataset. \textbf{Left:} window size $w$ affects pattern granularity (optimal: $w=3$); \textbf{Middle:} number of retrieved memories in linking stage $|L|$ (optimal: $|L|=5$); \textbf{Right:} number of memories used in ranking stage $K$ (optimal: $K=5$).}

    \label{fig:hyperparameter_study}
\end{figure*}
\subsubsection{Hyperparameter Analysis}
In Figure~\ref{fig:hyperparameter_study}, we present line plots illustrating the impact of three hyperparameters on the performance of our method. The first hyperparameter is the window size $w$, which defines a group of pattern items within a sliding window. This parameter plays a crucial role in determining the structure of extracted patterns. As shown in the figure, the best performance is achieved at $w=3$, which we therefore select as an appropriate setting in our experiments. For the number of top retrieved memories in the linking stage, denoted as $|\mathcal{L}|$, the optimal performance is achieved when $|\mathcal{L}| = 5$. The final hyperparameter is the top-$k$ memories used in the ranking stage. Increasing the number of memories from 3 to 5 in the memory pool leads to improved performance, yielding the best results. Overall, as the hyperparameter values increase, the performance tends to decrease and exhibit unstable behavior. This observation indicates that excessively large parameter settings introduce more noise than useful information, which can be detrimental to recommendation performance.

% \subsubsection{Link $L$ existing memories}

\subsubsection{Cold-start analysis}
To analyze the generalization ability of the memory pool, we conduct an experiment on a cold-start user dataset. In this setting, we evaluate our memory system on users who interact with only a few items (2–3 items). We compare AMEM4Rec with AMEM4Rec w/o ME, LLMRRank, and AgentCF. The results are reported in Table~\ref{tab:cold_start}. Our full AMEM4Rec model consistently outperforms all baseline methods, demonstrating its effectiveness in challenging cold-start scenarios with sparse user interactions. These results indicate that the patterns aggregated from cross-user behaviors provide robust and informative signals, enabling the model to maintain superior recommendation performance even under data sparsity.
% \begin{table}[!t]
% \centering
% \begin{tabular}{lccc}
% \toprule
% Method & Fashion &CDs and Vinyl& Video Games \\
% \midrule
% LLMRank& 0.5515& 0.4269& 0.5728\\
% AgentCF & 0.4105 & 0.3431 & 0.3730\\ 
% iAgent & 0.4240 & 0.4240 & 0.4666\\\hline
% AMEM4Rec w/o EM & 0.5621 & 0.4255& 0.6099 \\
% AMEM4Rec & 0.5723 & 0.4277& 0.6128\\
% \bottomrule
% \end{tabular}
% \caption{Cold-start user}
% \label{tab:cold_start}
% \end{table}
\begin{table}[!t]
\centering
\small
\setlength{\tabcolsep}{6pt}
\begin{tabular}{lccc}
\toprule
\textbf{Method} & \textbf{Fashion} & \textbf{CDs and Vinyl} & \textbf{Video Games} \\
\midrule
LLMRank
& 0.5515 & 0.4269 & 0.5728 \\

AgentCF
& 0.4105 & 0.3431 & 0.3730 \\ 

iAgent
& 0.4240 & 0.4240 & 0.4666 \\
\midrule
AMEM4Rec w/o EM
& 0.5621 & 0.4255 & 0.6099 \\

\textbf{AMEM4Rec}
& \textbf{0.5723} & \textbf{0.4277} & \textbf{0.6128} \\
\bottomrule
\end{tabular}
\caption{Performance comparison on cold-start users (2-3 interactions). AMEM4Rec 
shows strong generalization ability through cross-user memory patterns, 
reported in NDCG@10.}
\label{tab:cold_start}
\end{table}

\vspace{-1em}

% \subsubsection{Cross Domain}

\subsubsection{LLM Backbone Study}
To demonstrate the generalizability of AMEM4Rec, we evaluate its performance using different LLM backbones as the underlying agent model. Table~\ref{tab:llm_backbone} compares the full framework when powered by three representative models: Gemini 2.5 Flash API, Qwen2.5-7B-Instruct (open-source), and Llama-3-8B-Instruct (open-source). 

As shown in Table~\ref{tab:llm_backbone}, AMEM4Rec maintains strong and consistent performance across both closed-source and open-source LLM backbones. The results indicate that the core mechanism—memory creation, cross-user linking, and iterative evolution—effectively captures collaborative signals regardless of the underlying LLM backbone. Specifically, the stronger model (Gemini 2.5 Flash) yields higher results compared to Qwen2.5-7B-Instruct and Llama-3-8B-Instruct. This is reasonable given that the agent's tasks (semantic pattern extraction, linking decisions, and evolution reasoning) are complex and benefit from more advanced reasoning capabilities. 
% \section{Conclusion and Future}
\begin{table}[t]
\centering
\small
\setlength{\tabcolsep}{8pt}
\begin{tabular}{lcc} 
\toprule
\textbf{LLM Backbone} & \textbf{Video Games} & \textbf{MIND} \\
\midrule
Qwen2.5-7B-Instruct
& 0.3715 & 0.3240 \\

Llama-3-8B-Instruct
& 0.2532 & 0.3284 \\

\textbf{Gemini 2.5 Flash}
& \textbf{0.6108} & \textbf{0.4245} \\
\bottomrule
\end{tabular}
\caption{Performance comparison across different LLM backbones on Video Games and 
MIND datasets (NDCG@10).}
\label{tab:llm_backbone}
\end{table}

% \begin{table}[t]
% \centering

% \begin{tabular}{lcc} 
% \toprule
% LLM Backbone & Video Games & MIND \\
% \midrule
% Qwen2.5-7B-Instruct & 0.3715 & 0.3240\\
% Llama-3-8B-Instruct & 0.2532 & 0.3284 \\
% Gemini 2.5 Flash & 0.6108 & 0.4245 \\
% \bottomrule
% \end{tabular}
% \caption{Robustness across different LLM backbones. Results are reported in NDCG@10 on Video Games and MIND datasets.}
% \label{tab:llm_backbone}
% \end{table}

% \begin{acks}
% To Robert, for the bagels and explaining CMYK and color spaces.
% \end{acks}

\section{Conclusion and Future Work}

In this paper, we proposed AMEM4Rec, an agentic LLM-based recommender system that explicitly incorporates collaborative filtering signals through cross-user memory evolution. By creating abstract behavior memories from user histories, linking similar patterns across users, and iteratively evolving them, AMEM4Rec enables end-to-end collaborative learning without relying on any pretrained CF models. Extensive experiments on Amazon datasets (Fashion, Video Games, CDs and Vinyl) and the MIND news dataset demonstrate that AMEM4Rec consistently outperforms strong baselines, including prompt-based and generative LLM recommenders, as well as hybrid methods that inject CF knowledge. Ablation studies further confirm the critical role of memory evolution and link generation in aggregating collaborative signals and improving ranking quality, particularly in sparse interaction scenarios.

Despite these promising results, our approach still faces limitations. The performance heavily depends on the quality of LLM prompting and embedding models, and API costs remain a practical concern for large-scale deployment. In addition, while memory evolution effectively captures cross-user patterns, extremely sparse datasets may still limit the propagation of collaborative signals.

For future work, we aim to integrate purely textual memory representations with learnable parameters that can be directly optimized. By training these parameters using reinforcement learning \cite{wang2025mem,yan2025memory}, the agent can more intelligently control memory retrieval and evolution, leading to more adaptive and precise recommendations.

% First, integrating multi-modal inputs (e.g., images and text) into memory creation to support richer user behavior modeling. Second, exploring real-time memory evolution for dynamic recommendation scenarios. Third, scaling the framework to millions of users and evaluating its efficiency on distributed systems. Finally, we aim to combine AMEM4Rec with sequential recommendation models to better capture temporal dependencies. These extensions will further enhance the practicality and robustness of agentic LLM recommenders in real-world applications.
% %%
%% The next two lines define the bibliography style to be used, and
%% the bibliography file.
\bibliographystyle{ACM-Reference-Format}
\bibliography{ref}

%%
%% If your work has an appendix, this is the place to put it.
\appendix
% \section{Implementation Details}
\label{apendix:imp}
\section{Prompt Templates}
\label{apendix:prompt}

\lstset{
  basicstyle=\ttfamily\scriptsize,
  breaklines=true,
  breakatwhitespace=true,
  columns=fullflexible,
  keepspaces=true,
  showstringspaces=false
}

\begin{tcolorbox}[
  colback=lightgray,
  colframe=frameblue,
  arc=2mm,
  boxrule=0.8pt,
  left=6pt,
  right=6pt,
  top=6pt,
  bottom=6pt,
  listing engine=listings
]

\textbf{Prompt for Abstract Behavior Pattern Extraction}

\vspace{6pt}

\begin{lstlisting}


You are an expert analyst extracting abstract user behavior patterns
from Amazon shopping interactions.

Input (recent interactions):
{json.dumps(interaction_summary, indent=2)}

Rules (STRICT):
- Focus ONLY on category-level trends and general behavioral tendencies.
- NEVER use specific item names, brands, titles, franchises, or platforms
  in behavior_explanation.
- For pattern_description, mention specific items ONLY if they clearly
  represent a general sequence or co-occurrence
  (e.g., "console -> accessory").
- Keep both explanations VERY concise.

Extract:
1. behavior_explanation: 1-2 concise sentences describing stable,
   high-level user tendencies.
2. pattern_description: 1-2 concise sentences describing concrete
   interaction structure (sequence, repetition, co-purchase).

Return ONLY valid JSON:
{
  "behavior_explanation": "...",
  "pattern_description": "..."
}
\end{lstlisting}

\end{tcolorbox}

\begin{tcolorbox}[
  colback=lightgray,
  colframe=frameblue,
  arc=2mm,
  boxrule=0.8pt,
  left=6pt,
  right=6pt,
  top=6pt,
  bottom=6pt,
  listing engine=listings
]

\textbf{Prompt for Collaborative Pattern Linking}

\vspace{6pt}

\begin{lstlisting}
Determine if the new behavior pattern should be linked to past patterns
based on cross-user collaborative signals.

New Pattern:
- Behavior: {new_memory.behavior_explanation}
- Pattern: {new_memory.pattern_description}

Similar Past Patterns:
{json.dumps(nearest_info, indent=2)}

DECISION RULES:
- These patterns are already identified as similar
  (similarity >= low_threshold).
- Decide whether to UPDATE existing patterns or STORE the new pattern
  separately.
- UPDATE if: patterns share core categories, preferences, or behavioral
  trends that can be merged.
- STORE SEPARATELY if: despite high similarity, the new pattern reflects
  a genuinely different user segment.

Do NOT link only if patterns are completely unrelated.

Return JSON:
{
  "should_link": true/false,
  "linked_thought_ids": [list of IDs],
  "reasoning": "1-2 concise sentences explaining the decision"
}
Ensure reasoning is concise and specific.
\end{lstlisting}

\end{tcolorbox}

\begin{tcolorbox}[
  colback=lightgray,
  colframe=frameblue,
  arc=2mm,
  boxrule=0.8pt,
  left=6pt,
  right=6pt,
  top=6pt,
  bottom=6pt,
  listing engine=listings
]

\textbf{Prompt for Collaborative Memory Evolution}

\vspace{6pt}

\begin{lstlisting}
You are a collaborative memory evolution expert.
Your goal is to reinforce shared cross-user patterns with minimal,
concise changes.

New Pattern:
- Behavior: {new_memory.behavior_explanation}
- Pattern: {new_memory.pattern_description}

Candidate Memories (with evolution count):
{json.dumps(mem_info, indent=2)}

Rules:
- Update only if the new pattern meaningfully strengthens or refines
  a shared category, preference, or sequence.
- Keep updated text VERY concise.
- If no meaningful improvement, keep the original text.

Return JSON:
{
  "should_evolve": true/false,
  "updates": [
    {
      "thought_id": ID,
      "behavior_explanation": "updated text or null",
      "pattern_description": "updated text or null",
      "reasoning": "1 short sentence explaining the update"
    }
  ]
}
\end{lstlisting}

\end{tcolorbox}

\begin{tcolorbox}[
  colback=lightgray,
  colframe=frameblue,
  arc=2mm,
  boxrule=0.8pt,
  left=6pt,
  right=6pt,
  top=6pt,
  bottom=6pt,
  listing engine=listings
]

\textbf{Prompt for Ranking}

\vspace{6pt}

\begin{lstlisting}
You are ranking candidate items for a user based on their shopping history.

Inputs:

User Recent History (prioritize most recent):
{json.dumps(user_profile, indent=2)}

Collaborative Memory Insights (cross-user behavior patterns):
{json.dumps(memory_thoughts, indent=2)}

Candidate Items:
{json.dumps(candidate_info, indent=2)}

Ranking Rules:
- Prioritize items matching the user's most recent interactions and memory
  insights (e.g., if memory shows "prefers action games", rank action games higher).
- Ensure category consistency with history and memory patterns.
- For similar items, prefer those aligning with cross-user trends in memory
  (e.g., popular items in the same category).
- If history or memory is limited, favor items with broader category relevance.

Output Requirements:
- Rank ALL {len(candidate_items)} candidate items.
- Return ONLY valid JSON.
- Reasoning must be concise and avoid repeating item titles or categories.

JSON Format:
{
  "ranked_item_ids": ["item_id1", "item_id2", "..."],
  "reasoning": "1 sentence explaining ranking logic, focusing on history and memory alignment"
}
\end{lstlisting}

\end{tcolorbox}

% \section{Research Methods}

% \subsection{Part One}

% Lorem ipsum dolor sit amet, consectetur adipiscing elit. Morbi
% malesuada, quam in pulvinar varius, metus nunc fermentum urna, id
% sollicitudin purus odio sit amet enim. Aliquam ullamcorper eu ipsum
% vel mollis. Curabitur quis dictum nisl. Phasellus vel semper risus, et
% lacinia dolor. Integer ultricies commodo sem nec semper.

% \subsection{Part Two}

% Etiam commodo feugiat nisl pulvinar pellentesque. Etiam auctor sodales
% ligula, non varius nibh pulvinar semper. Suspendisse nec lectus non
% ipsum convallis congue hendrerit vitae sapien. Donec at laoreet
% eros. Vivamus non purus placerat, scelerisque diam eu, cursus
% ante. Etiam aliquam tortor auctor efficitur mattis.

% \section{Online Resources}

% Nam id fermentum dui. Suspendisse sagittis tortor a nulla mollis, in
% pulvinar ex pretium. Sed interdum orci quis metus euismod, et sagittis
% enim maximus. Vestibulum gravida massa ut felis suscipit
% congue. Quisque mattis elit a risus ultrices commodo venenatis eget
% dui. Etiam sagittis eleifend elementum.

% Nam interdum magna at lectus dignissim, ac dignissim lorem
% rhoncus. Maecenas eu arcu ac neque placerat aliquam. Nunc pulvinar
% massa et mattis lacinia.

\end{document}